\documentclass{article}
\usepackage{bigsky2007}
\usepackage{floatflt,epsfig}
\usepackage{subfigure}
\usepackage{xspace}

\frompage{000} \topage{000}                                              

\newcommand{\pt}{\ensuremath{p_{T}}\xspace}

\newcommand{\pp}{\ensuremath{p+p}\xspace}
\newcommand{\PbPb}{\ensuremath{Pb+Pb}\xspace}

\newcommand{\pizero}{\ensuremath{\pi^{0}}\xspace}

\newcommand{\sqsPP}{\ensuremath{\sqrt{s}=14}\xspace}
\newcommand{\sqsPbPb}{\ensuremath{\sqrt{s}=5.5}\xspace}
\newcommand{\sqsPPb}{\ensuremath{\sqrt{s}=8.8}\xspace}
\newcommand{\Lxy}{\ensuremath{L_{xy}}\xspace}

\def\rightmark{Reconstructing displaced Bottom vertices}
\def\leftmark{M.Heinz (Yale)}

\title{Reconstructing Bottom mesons using displaced vertices from
semi-leptonic decays }
\authors{
{Mark T. Heinz (for the ALICE collaboration)
}\\[2.812mm]
{\normalsize Yale University, WNSL, 272 Whitney Avenue,CT 06517, New Haven, USA\\[0.2ex]
E-mail: \texttt{mark.heinz@yale.edu} }}

\abstract{Precise determination of heavy flavor production
cross-sections at LHC energies will be of primary importance. The
produced heavy quarks are expected to be sensitive probes of parton
energy loss in the medium formed in heavy-ion collisions. Through
the measurement of charm and bottom suppression in \PbPb with
respect to \pp, we hope to obtain insight into the color-charge and
quark mass dependence of the energy loss mechanism. The ALICE
experiment with its large acceptance is well suited to investigate
the intermediate transverse momentum spectrum of heavy flavor mesons
where these energy loss effects are expected to be visible. ALICE
has very good electron PID capabilities over a large kinematical
range using Time Projection Chamber (TPC), Transition Radiation
Detector (TRD) and the Electromagnetic Calorimeter (EMCal). In
addition the EMCal, to be installed for the \PbPb runs, is planned
to allow efficient triggering on high-\pt jets. We first introduce
the EMCal project and an overview of detector specifications. Then
we introduce a method developed to select preferentially electrons
from heavy flavor decays by reconstructing displaced secondary
vertices. The strategy is to reconstruct displaced vertices from
semi-leptonic heavy flavor meson decays using the excellent spatial
resolution of the Inner Tracking System (ITS). We show preliminary
results of an efficiency study from charm vs. bottom vertices.}


\begin{document}

\maketitle

\setcounter{page}{1}

\section{Introduction}\label{intro}

The ALICE experiment is a dedicated heavy-ion experiment at the
Large Hadron Collider (LHC) at CERN in Geneva, Switzerland. The LHC
will provide for \pp ~collisions at \sqsPP TeV as well as \PbPb and
$p+Pb$ heavy-ion collisions at \sqsPbPb TeV respectively \sqsPPb
TeV.

The EMCal in ALICE will be ideally suited for the study of QCD
matter at high temperatures by probing the hard parton scattering
processes. Once integrated into ALICE the EMCal will allow
triggering on high-\pt particles such as photons, \pizero and
electrons. In particular, one of the goals is to study fragmentation
functions by tagging jets with a specific quark content, e.g. heavy
flavor or b-jets.

The importance of identifying heavy flavor jets at ALICE is twofold.
First, from a basic perturbative QCD standpoint the precise
determination of transverse momentum spectra for charm and bottom
hadrons is of primary importance in understanding the production
mechanisms of heavy quarks. The current theoretical fixed-order
next-to-leading log calculations (FONLL) still carry large
uncertainties due to poorly known heavy quark masses and differences
in parton distribution functions \cite{AliceNote:Dainese}. Second,
the propagation and energy loss of heavy quarks in the dense medium
created in heavy-ion collisions is of great interest since these
`hard' probes will help determine basic properties of the medium.
Recent measurements at RHIC have shown that the non-photonic
electrons, expected to mainly originate from heavy flavor decays,
are strongly suppressed, possibly indicating that the energy loss of
charm and bottom quarks is larger than predicted by theoretical
calculations using radiative energy loss
\cite{STAR:eSuppression,BDMPS,DGLV}. In order to understand this
discrepancy we need to measure charm and bottom energy loss
separately at the LHC. In section \ref{method} of this paper a
method is described for measuring B-mesons via their semi-leptonic
decay channel.
\begin{figure}[h]
\centering \mbox{
\subfigure[]{\includegraphics[width=6cm]{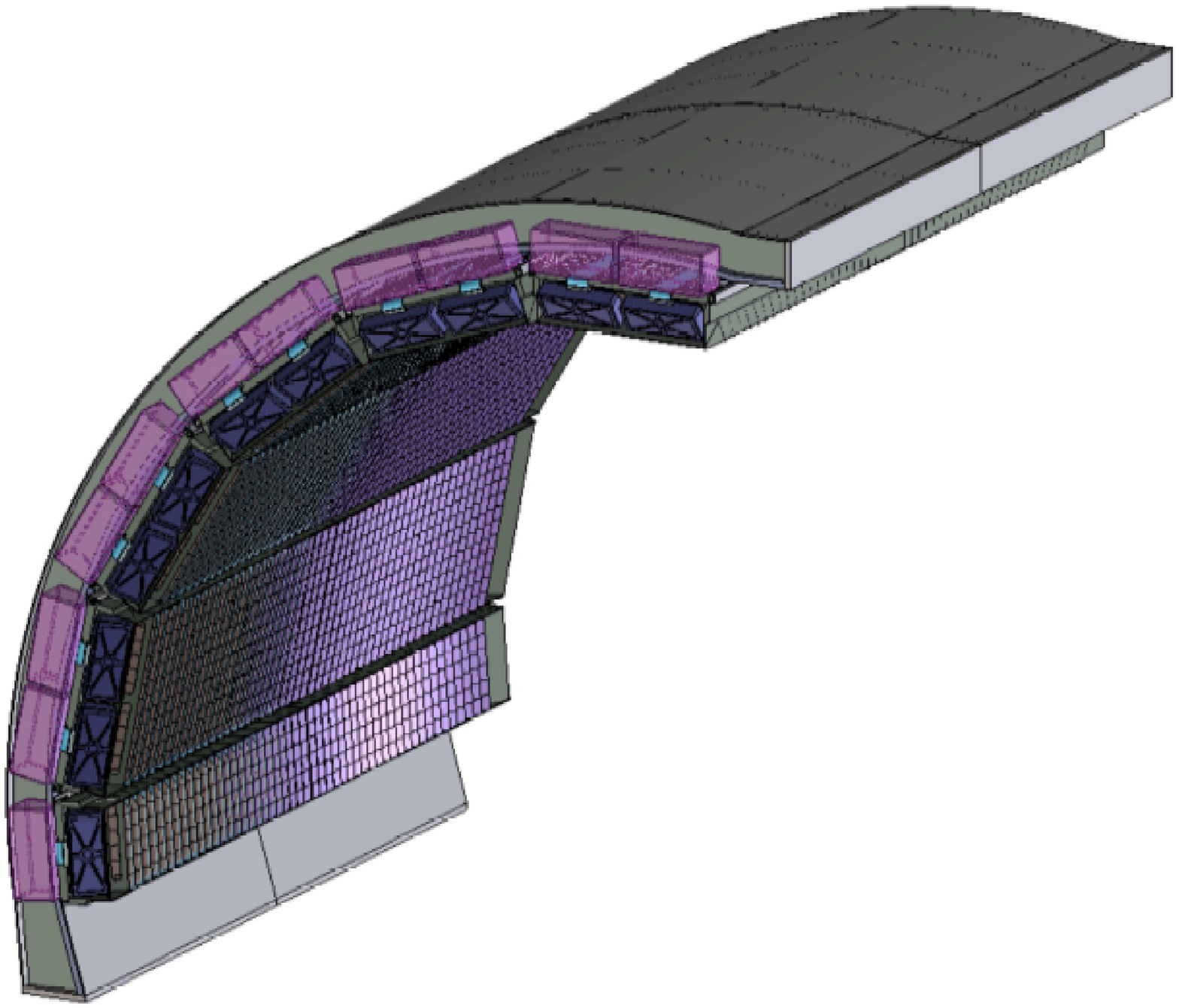}} \quad
\subfigure[]{
\includegraphics[width=6cm]{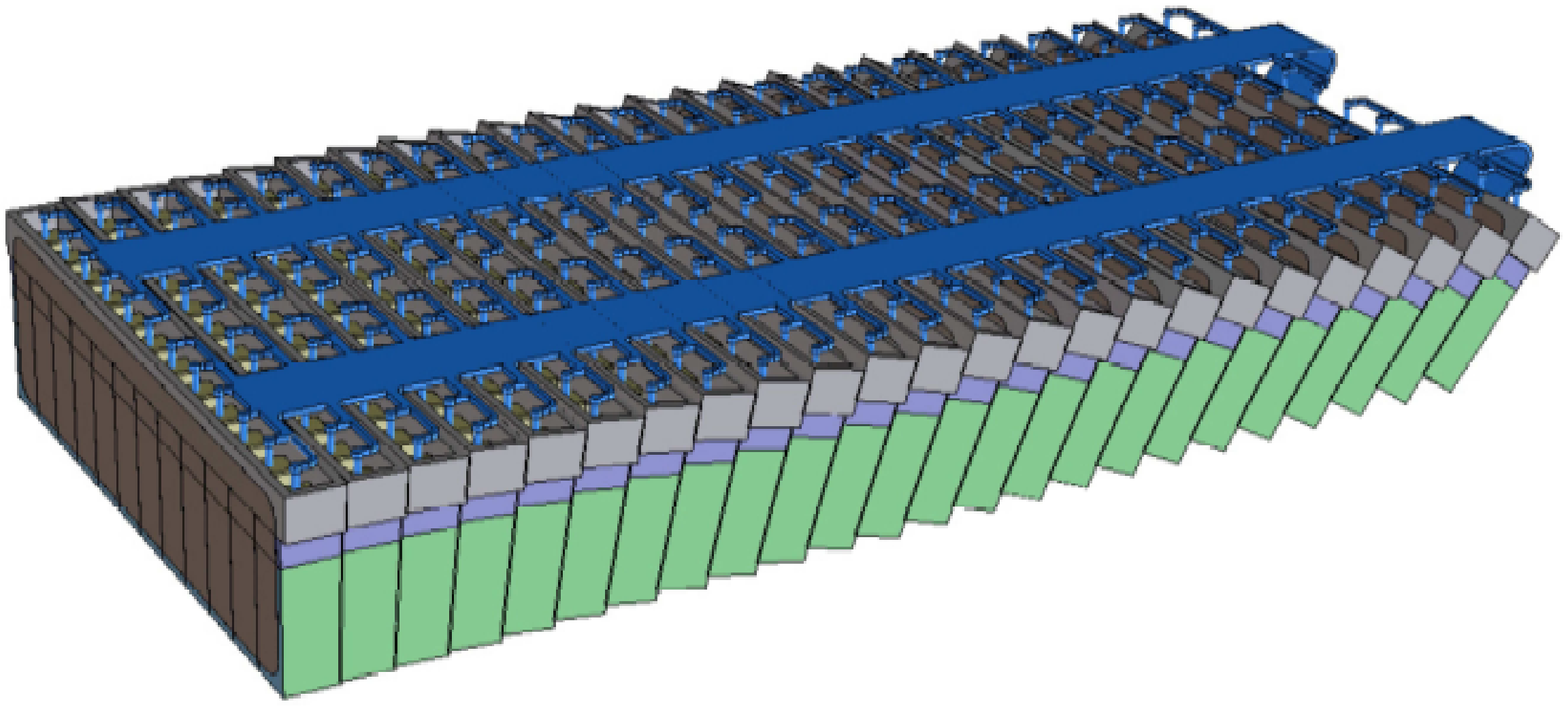}} }
\caption{(a) Drawing of complete EMCal with 11 supermodules mounted
on the support structure, (b) Drawing of supermodule consisting of
12 x 24 modules }\label{fig:emcal}
\end{figure}

\section{ALICE EMCal}\label{emcal}

The ALICE EMCal is proposed to enhance ALICE's high momentum
particle measurements and, in particular, improve its capabilities
for the jet quenching measurements. It will also allow
implementation of high-level jet-triggers which should significantly
improve both the statistics and the energy resolution of jets.
Further, the EMCal will complement ALICE's particle identification
(PID) capabilities for high momentum photons, neutral pions and
electrons.

The project is jointly funded by US, French and Italian institutes.
The ALICE-USA collaboration counts 12 member institutions involved
in the EMCal project. After initial funding approval, expected in
September 2007, the construction will commence and first
supermodules should be delivered to CERN in early 2009. The
installation of all 11 supermodules, to be assembled in the US and
France, is foreseen to be completed in 2011.

The conceptual design of the EMCal is based on the Shashlik
technology as implemented for example in the PHENIX experiment at
RHIC \cite{Phenix:Ecal}. Figure \ref{fig:emcal}(a) shows the EMCal
supermodules mounted in the installed position on their support
structure. They each span about 20 degrees in azimuth and about 0.7
units of pseudorapidity. There are 10 full size and 2 half-sized
supermodules in the full detector acceptance. Figure
\ref{fig:emcal}(b) shows a detail of the supermodule which is the
basic structural unit of the calorimeter and consists of 288 modules
(12x24). The full detector spans $\eta = [-0.7,0.7]$ with an
azimuthal acceptance of $\Delta\phi = 110 \deg$.

The chosen technology is that of a layered Pb-scintillator sampling
calorimeter with a longitudinal pitch of 1.44mm Pb and 1.76 mm
scintillator with longitudinal wavelength shifting fibre light
collection (Shashlik). The EMCal is segmented into 12672 towers,
grouped into 2x2 modules, each of which is approximately projective
to the interaction vertex in $\eta$ and $\phi$. The front face
dimensions of the towers are 6x6 $cm^{2}$ (Moliere radius $\sim$
2cm) resulting in individual tower acceptance of
$\Delta\eta\times\Delta\phi = 0.014 \times 0.014$. The energy
resolution as determined from GEANT simulation is $12\% / \sqrt{E} +
2\%$. A more detailed description of the detector and its expected
performance can be found in \cite{TP:Emcal}.

The EMCal will improve the PID coverage for photon/\pizero
discrimination through cluster shape analysis. Further
electron/hadron separation can be achieved by the energy/momentum
method (chapter 7.4 \cite{TP:Emcal}).

\begin{figure}[h]
\centering \mbox{
\subfigure[]{\includegraphics[width=6cm]{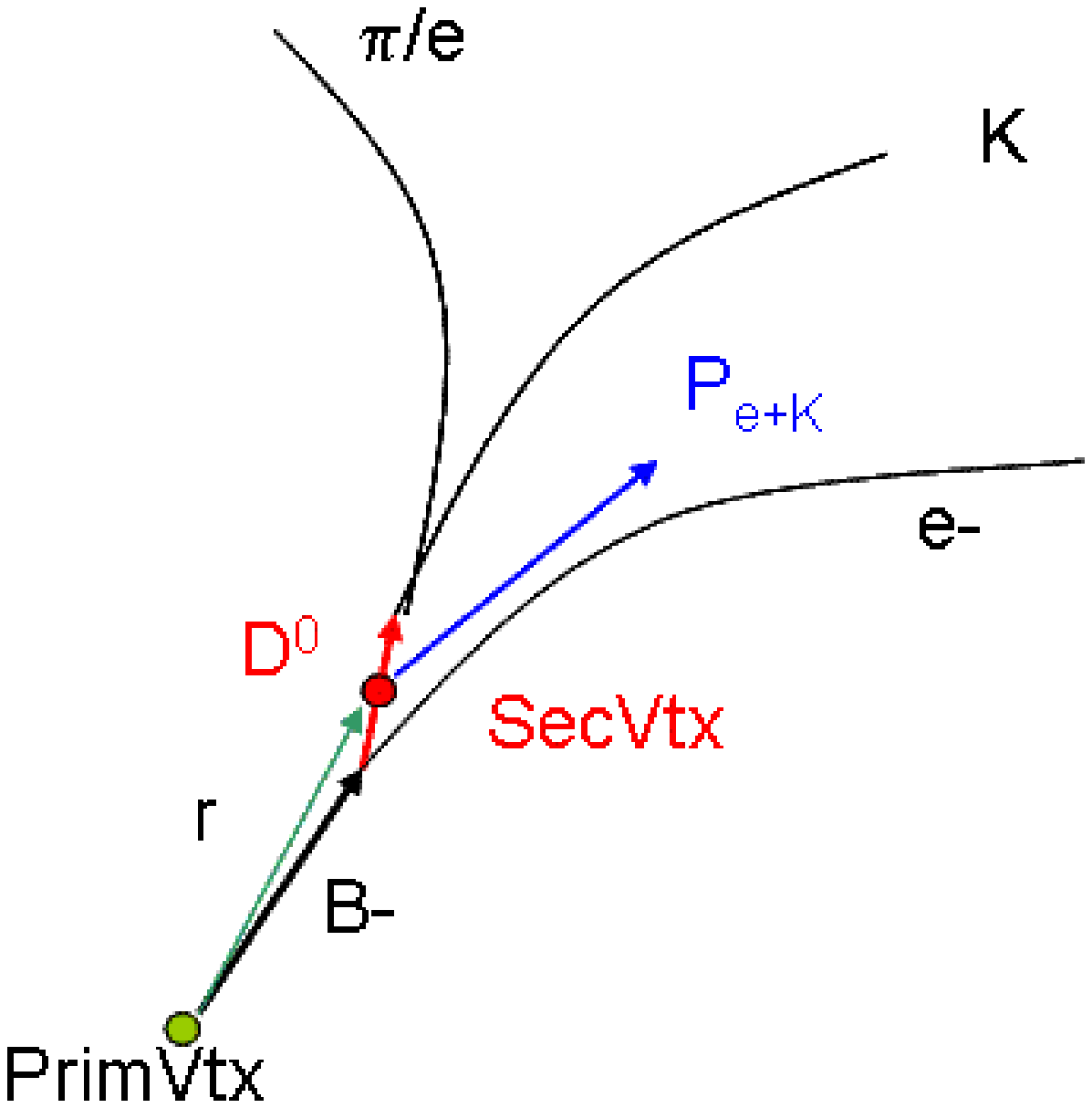}} \quad
\subfigure[]{\includegraphics[width=6cm]{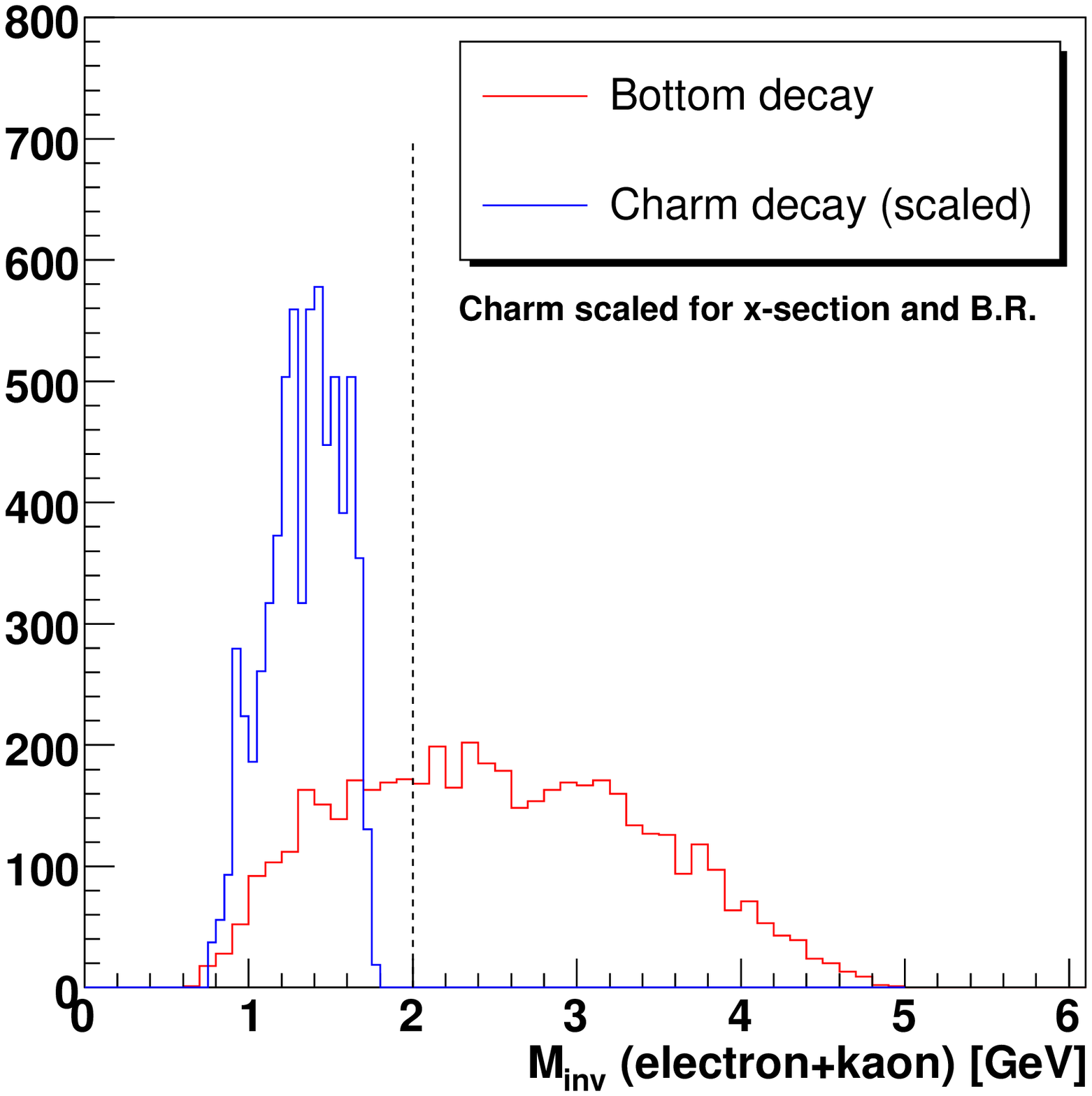}}
} \caption{(a) Scheme of the preferred $B^-$ decay (only charged
tracks shown), (b) Invariant mass (assuming the kaon mass for the
hadron) before geometrical cuts of electron+hadron pair for charm vs
bottom decays. The charm curve (blue) was scaled for the expected
ratio (pQCD) in cross-section (factor 22.0) and semi-leptonic
branching ratios (factor 0.87). The vertical line indicates the
chosen cut value for this study.} \label{fig:Bdecay}
\end{figure}

\section{Reconstructing B-jets via displaced vertices}\label{method}

Several methods for identifying heavy flavor decays via the
semi-leptonic channel have been discussed in the ALICE PPR Vol 2
\cite{ALICE:PPR}. The method described herein is complementary to
these methods and is especially well suited for high-\pt electrons
which can be identified with a combination of ALICE detectors as
mentioned earlier.

The method was first utilized in the CDF experiment to identify
bottom via their muon decay channels \cite{CDF:Method}. It relies on
the fact that the approximate position of the semi-leptonic B-decay
vertex can be reconstructed using two tracks, one of which is the
lepton. B-meson decays have a $c\tau \sim500\mu$m making it easier
to resolve this secondary vertex than for charm ($c\tau
100-300\mu$m). At the B-decay vertex a lepton and a charmed meson
($D^{0},D^{\pm},D^{*}$) is often produced and subsequently decays
into charged hadrons. The method starts off by finding high-\pt
electrons and combining them with all charged particles within a
wide jet-cone of radius $dR^{2} = \Delta\eta^{2}+\Delta\phi^{2} <
1.5$. As shown in figure \ref{fig:Bdecay}(a) using any two tracks
(electron + kaon/pion) an approximate secondary vertex can be
calculated from their distance of closest approach (DCA). For this
study we assume that the charm decay length is not resolved and
therefore the length of the vector r (see figure
\ref{fig:Bdecay}(a)) approximates the displaced vertex. There are
other decays in which the electron originates from a D or the
associated hadron is produced directly at the B-vertex. Using r and
the momentum-sum of the electron and pion/kaon $p_{e+K/\pi}$ (which
carries most of the B-meson momentum) a quantity called the signed
decay-length, $L_{xy}$ is defined in the bending plane according to
equation \ref{Eq:Lxy}.

\begin{equation}
L_{xy} = \frac{r \cdot p_{e+K}}{|p_{e+K}|} = |r| \cdot cos(\theta)
\label{Eq:Lxy}
\end{equation}

For `real' decays this quantity naturally has to be larger than 0
since the angle $\theta$ between r and $p_{e+K}$ is less than 90
degrees. Background pairs, on the other hand, are assumed to be
evenly distributed over all values of $\theta$. The background
trigger electrons for this method will mainly consist of electrons
from photonic decays (conversions, $\pizero \rightarrow \gamma
\gamma$), \pizero dalitz decays, and electrons from semi-leptonic
decays of primary charm hadrons.

\begin{figure}[ht]
  \includegraphics[width=\textwidth]{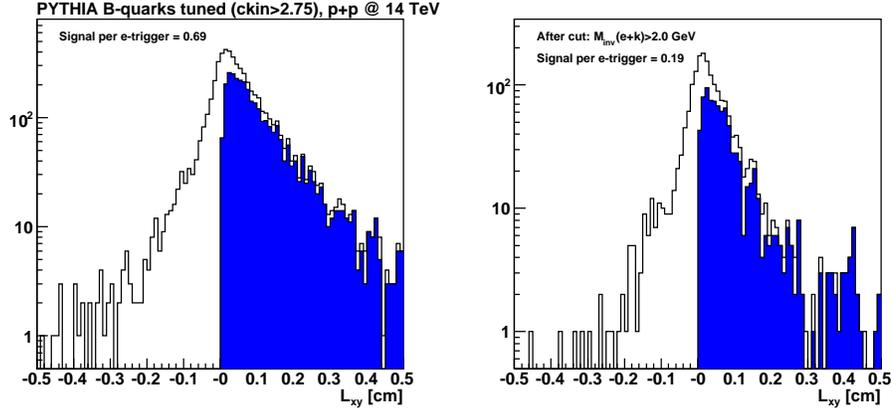}
  \caption{$L_{xy}$ of bottom sample before (left) and after (right) invariant mass cut.
  The blue histograms are signal distributions after subtraction of the negative \Lxy part.}
  \label{SignDcaBottom}
\end{figure}

Since the charm cross-section is about 20 times larger (at 14 TeV)
than the bottom cross-section it is necessary to find an efficient
way of eliminating the charm signal in our sample. As shown in
figure \ref{fig:Bdecay}(b) this can be done by applying a lower cut
on the invariant mass of the electron and the hadron, under the
assumption that the hadron is a kaon (using the pion mass works
equally). For charm decays this quantity is approximately bounded by
the charm meson mass, i.e. $\sim 1.9$ GeV. For bottom decays it is
bounded by the bottom meson mass, 5.279 GeV, which is much higher.

\section{Simulation results}\label{results}

We started by applying this method to ALICE PYTHIA simulations in
order to get a first estimate of bottom signal efficiency and charm
rejection rate. Since the simulation code is still subject to
frequent modifications it is useful to state that these results are
based on the ALIROOT HEAD version of November 2006 \cite{Aliroot}.
These simulation do not yet include all detector material and
misalignment they are to be considered idealistic.

In order to obtain initial efficiency estimates of this method three
simulation samples were produced:
\renewcommand{\theenumi}{\alph{enumi}}
\begin{enumerate}
\item{PYTHIA (tuned to match NLO) bottom sample ($b-\bar{b}$) from 14 TeV \pp collisions with forced semi-leptonic decay. 40k events.}
\item{PYTHIA (tuned to match NLO) charm sample ($c-\bar{c}$) from 14 TeV \pp collisions with forced semi-leptonic decay. 40k events.}
\item{PYTHIA minbias sample (suppressed B,D decays) from 14 TeV \pp collisions. 100k events. }
\end{enumerate}

The first sample (a) allowed signal efficiencies for different
particle and vertex selection cuts to be determined. The second (b)
and third sample (c) were used to estimate the backgrounds from
charm decays respectively photonic conversions and \pizero decays in
\pp collisions. For this first study we applied the following
electron, hadron and vertex selection cuts. To reduce the computing
time for simulations we here only used electron PID from the TPC,
but will include TRD and EMCal in the future.

\begin{itemize}
\item{Electrons: TPC-PID probability $>80\%$ (dE/dx, as implemented in class AliPID, Section 5.4.6 of \cite{ALICE:PPR}), $p_{T}> $1.5 GeV/c, 4 or more ITS-Hits, impact(xy)$< 0.5$ cm, impact(z)$<1.0$ cm }
\item{Hadrons: 4 or more ITS-Hits, impact(xy)$< 0.5$ cm, impact(z)$<1.0$ cm}
\item{Vertex: DCA between electron-hadrons$< 1$ mm, distance to primary vertex $< 1$ cm, }
\end{itemize}

\begin{figure}[h]
  \includegraphics[width=\textwidth]{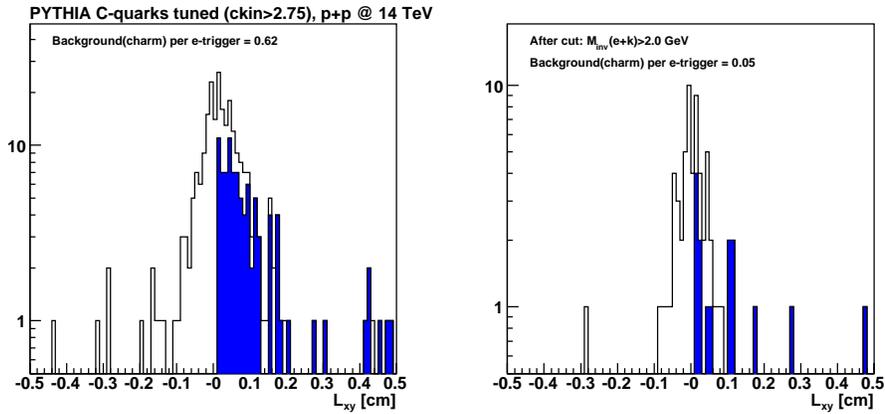}\\
  \caption{$L_{xy}$ of Charm sample before (left) and after (right) invariant mass cut}
  \label{SignDcaCharm}
\end{figure}

Using the above selection cuts we have analyzed the 3 simulation
samples and obtained signed decay-length distributions. Figure
\ref{SignDcaBottom} shows the result of the pure bottom production
(sample a). The signal (blue area) is defined as all positive
entries ($S+B: \Lxy^{pos} > 0 $) minus all negative entries ($B:
\Lxy^{neg} < 0$). The total signal is normalized by the number of
electron triggers. For the pure B-sample the efficiency with these
cuts is $\sim70\%$, whereas we reject $\sim30\%$ of the signal due
to other B-decay topologies and resolution effects. Figure
\ref{SignDcaBottom}(b) shows the signed decay-length distribution
after applying the invariant mass cut.

In figure \ref{SignDcaCharm} we have applied the same method to a
comparable sample of charm mesons. Although the number of trigger
electrons passing our \pt cut is lower due to the different
kinematics, the signal per electron trigger is not very different,
i.e. 62\%. This invariant mass cut reduces our bottom-efficiency to
$\sim20\%$ and our charm-efficiency to $\sim5\%$. According to pQCD,
the decay electron yield from charm and bottom at 2-3 GeV/c is of
comparable magnitude. By imposing an even tighter cut this charm
contamination can be reduced even more.

\begin{figure}[h]
\centering
\includegraphics[width=\textwidth]{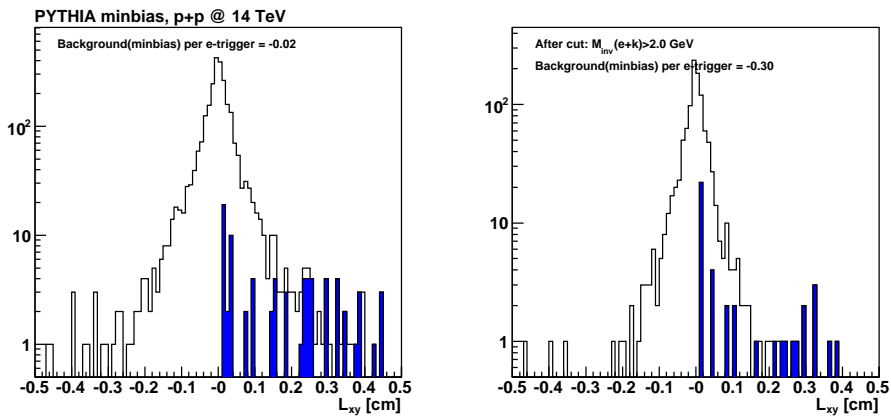}
\caption{\Lxy of Minbias PYTHIA sample before (left) and after
(right) invariant mass cut} \label{pythia}
\end{figure}

Finally, figure \ref{pythia} shows the signal for the PYTHIA Minbias
sample containing no B-decays with and without invariant mass cuts.
Clearly the statistics is not sufficient to make any quantitative
statements, however using these cuts the signal from this sample is
negative, due to the large contribution from photonic conversion
decays. The goal is to find optimal cuts to reduce as much as
possible these contributions. These studies are still ongoing.

\section{Summary and Outlook}\label{summary}

A method for identifying heavy flavor jets is presented based on the
reconstruction of displaced vertices from semi-leptonic decays. This
method can distinguish between bottom and charm contributions to the
non-photonic electron spectra. First simulation results in \pp
collisions indicate an efficiency for reconstructing B-vertices of
$\sim20\%$. The ALICE EMCal will allow for high-level electron
triggering and complement electron PID capabilities of the TPC and
TRD.

In the near future more realistic simulations will be produced to
fully investigate the potential using also TRD and EMCal electron
PID. Next, the backgrounds from conversions and combinatorics in \pp
and heavy ion events need to be studied. Finally, possible effects
of misalignment in the ITS also need to be simulated.

\vfill\eject
\end{document}